\documentclass[prl,aps,superscriptaddress,showpacs,twocolumn,reprint,noshowkeys,amssymb,amsmath,amsfonts]{revtex4}

\usepackage{theorem}
\usepackage{dcolumn}
\usepackage{longtable}
\usepackage{bm}
\usepackage{graphicx}

\newcommand{\SSS}{\scriptscriptstyle}
\newcommand{\DS}{\displaystyle}

\newcommand{\Dd}{{\rm d}}

\newcommand{\Ii}{{\rm i}}

\newcommand{\IM}{{\mathrm{Im}}}

\newcommand{\kappaD}{\kappa_{{\SSS\text{D}}}}
\newcommand{\BarEpsM}{\bar{\epsilon}_{\SSS\text{M}}}

\newcommand{\kappaM}{\kappa_{{\SSS\text{M}}}}

\newcommand{\dperp}{d_{\SSS\perp}}
\newcommand{\dpara}{d_{\SSS\parallel}}

\newcommand{\EpsB}{\epsilon_{\SSS\text{B}}}

\newcommand{\EpsM}{\epsilon_{{\SSS\text{M}}}}

\newcommand{\rB}{a_{0}}

\newcommand{\EpsD}{\epsilon_{\SSS\text{D}}}

\newcommand{\EpsDO}{\epsilon_{\SSS\text{D}0}}

\newcommand{\AlphaD}{\alpha_{{\SSS\textrm{D}}}}

\newcommand{\OmegaP}{\omega_{\text{p}}}

\newcommand{\GammaP}{\gamma_{\text{p}}}

\newcommand{\muF}{\mu_{\SSS\text{F}}}

\newcommand{\Vxc}{V_{\text{xc}}}

\begin{document}

\title{Quantum-Spillover-Enhanced Surface-Plasmonic Absorption at the Interface of Silver and High-Index Dielectrics}

\author{Dafei Jin}
\author{Qing Hu}
\affiliation{Department of Mechanical Engineering, Massachusetts Institute of Technology, Cambridge, Massachusetts 02139, USA}
\author{Daniel Neuhauser}
\affiliation{Department of Chemistry and Biochemistry, University of California at Los Angeles, Los Angeles, California 90095, USA}
\author{Felix von Cube}
\affiliation{School of Engineering and Applied Sciences, Harvard University, Cambridge, Massachusetts 02138, USA}
\author{Yingyi Yang}
\affiliation{Department of Mechanical Engineering, Massachusetts Institute of Technology, Cambridge, Massachusetts 02139, USA}
\author{Ritesh Sachan}
\affiliation{Materials Science and Technology Division, Oak Ridge National Laboratory, Oak Ridge, Tennessee 37831, USA}
\author{Ting S. Luk}
\affiliation{Sandia National Laboratory, Albuquerque, New Mexico 87123, USA}
\author{David C. Bell}
\affiliation{School of Engineering and Applied Sciences, Harvard University, Cambridge, Massachusetts 02138, USA}
\author{Nicholas X. Fang}\email{nicfang@mit.edu}
\affiliation{Department of Mechanical Engineering, Massachusetts Institute of Technology, Cambridge, Massachusetts 02139, USA}
\date{\today}

\begin{abstract}

We demonstrate an unexpectedly strong surface-plasmonic absorption at the interface of silver and high-index dielectrics based on electron and photon spectroscopy. The measured bandwidth and intensity of absorption deviate significantly from the classical theory. Our density-functional calculation well predicts the occurrence of this phenomenon. It reveals that due to the low metal-to-dielectric work function at such interfaces, conduction electrons can display a drastic quantum spillover, causing the interfacial electron-hole pair production to dominate the decay of surface plasmons. This finding can be of fundamental importance in understanding and designing quantum nano-plasmonic devices that utilize noble metals and high-index dielectrics.

\end{abstract}


\maketitle

\pretolerance=8000
\tolerance=2000

Surface plasmons (SPs),  collective oscillations of conduction electrons at a metal-dielectric interface, have attracted interest for several decades \cite{RaetherBook,LiebschBook,MaierBook}. Nanomaterials that strongly absorb visible light through plasmonic effects could be very important for solar-energy devices \cite{LinicNM2011,ClaveroNP2014,LiuNL2010,AydinNC2011,CuiNL2012}. It is normally assumed that classical theory, with prescribed frequency-dependent bulk permittivities, reliably captures the SP properties, while quantum effects, despite their academic interest \cite{SavageNature2012,EstebanNatCommun2012,ToscanoNatCommun2015,McMahonPRL2009,WienerNL2012,YanPRL2015}, usually have negligible effect in practical systems.

Here we show that contrary to conventional wisdom, quantum effects can play a crucial role for SPs at the interface of silver (Ag) and (practically any) high-index dielectrics. Our density-functional calculation and spectroscopic measurement indicate the existence of remarkable non-classical plasmonic absorption in such systems. Interfacial electron-hole (e-h) pair production \cite{LiebschBook,
FeibelmanPSS1982,PerssonPRB1985,KempaPRB1986,RoccaSSR1995,LiNJP2013} can become the predominant decay channel of SPs, exceeding the ordinary phonon scattering (the Drude loss) or interband transitions (the dielectric loss). The quantum origin of such plasmonic absorption has been largely overlooked in nano-plasmonics research.

\begin{figure}[htb]
\centerline{\includegraphics[scale=0.8]{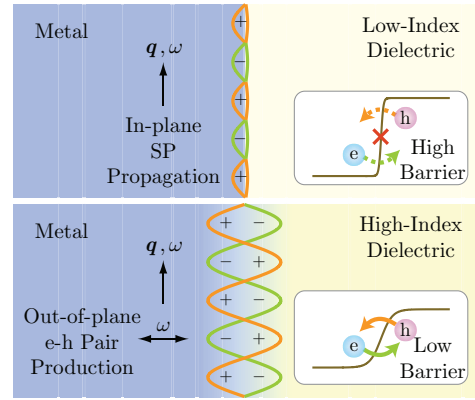}}
\caption{Schematics for the coupling of surface plasmons and interfacial electron-hole pairs at the interface of a metal and a dielectric.  The e-h pair production can be viewed  as dipole transitions, across the low barrier of the high-index dielectric,  driven by the out-of-plane electric field of surface plasmons.}
\label{FigSketchPairs}
\end{figure}

High-index dielectrics have been long considered as superior gate insulators in nano-electronics \cite{RobertsonRPP2006}. In the recent years, they have also attracted much interest in nanophotonics on the conversion of SPs to hot electrons \cite{ClaveroNP2014,BrongersmaNatMat2014,LeeNL2011,LeeJPCC2014,DuCheneAngew2014}. However, very limited attention has been paid to their quantum electronic properties that can reversely modify the SP response. The uniqueness of high-index dielectrics, compared with low-index dielectrics lies in their large electron affinity, high static permittivity, and thereby a much lowered work function to the conduction electrons in metal. This allows the electrons to undergo a deep quantum spillover into the high-index dielectrics extending beyond the Thomas-Fermi screening length. The in-plane SP propagation can strongly couple with the out-of-plane e-h pair production, as illustrated in Fig.~\ref{FigSketchPairs}. This process is insignificant at a metal-low-index interface \cite{LiebschBook,FeibelmanPSS1982,PerssonPRB1985} due to the high barrier and thus insufficient quantum spillover. But it is remarkable at a metal-high-index interface and can lead to intensified energy dissipation and widened absorption spectrum.

\begin{figure}[htb]
\centerline{\includegraphics[scale=0.8]{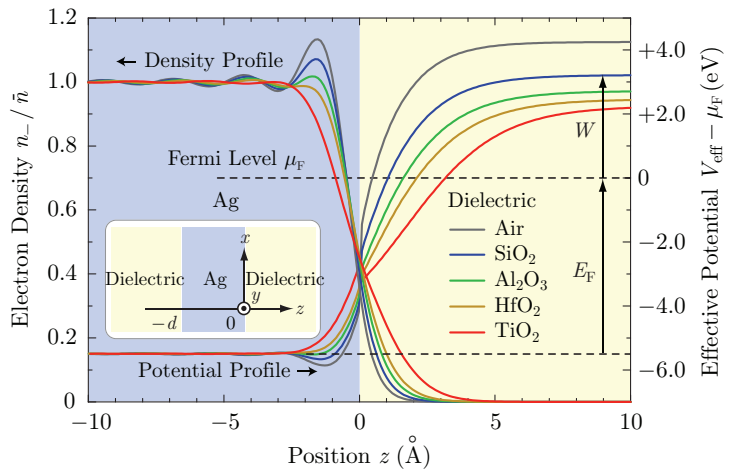}}
\caption{Calculated ground-state electron-density profiles, and effective-potential profiles with respect to the Fermi level $\mu_{\SSS\text{F}}$.}
\label{FigGroundState}
\end{figure}

Our theoretical study utilizes a generalized jellium density-functional model that incorporates the crucial properties of Ag and dielectrics. A uniform Ag slab with a thickness $d=100~\rB$ ($\rB$ is the Bohr radius) along $z$ is clamped by a dielectric on both sides (see the inset of Fig.~\ref{FigGroundState}). The conduction electrons are governed by the Kohn-Sham and generalized Poisson equations \cite{LangPRB1970,LangPRB1971,LiebschPRL1993,YuanPRB2006,GaoJCP2013},
\begin{align}
\left\{-\frac{\hbar^2}{2m} \frac{\Dd^2}{\Dd z^2} + V_{\SSS\text{eff}}[n_{\SSS-}(z)] \right\} \varphi_\nu(z) = \varepsilon_\nu  \varphi_\nu(z) ,\quad\\
\frac{\Dd}{\Dd z} \left\{ \epsilon (z;\omega)|_{\omega\rightarrow 0} \frac{\Dd}{\Dd z} \mathit{\Phi} (z) \right\} = 4\pi e \left\{ n_{\SSS-}(z)-n_{\SSS+}(z) \right\}.
\end{align}
$\varphi_\nu(z)$ and $\varepsilon_\nu$ are the eigenfunctions and eigenvalues. $n_{\SSS-}(z)$ is the electron number density and is a sum of $|\varphi_\nu(z)|^2$ over the occupied orbitals up to the Fermi level $\muF$ \cite{YuanPRB2006,GaoJCP2013}. $n_{\SSS+}(z)$ is the space-dependent positive-jellium density; $\epsilon(z;\omega)$ is the space- and frequency-dependent background permittivity that accounts for the screening from valance electrons.  The effective potential reads
\begin{align}
V_{\SSS\text{eff}}[n_{\SSS-}(z)] = -e \mathit{\Phi}(z) + \Vxc[n_{\SSS-}(z)] + \alpha(z) ,
\end{align}
in which $\Vxc[n_{\SSS-}(z)]$ is the exchange-correlation potential, and $\alpha(z)$ is the space-dependent electron affinity in a flat-band picture \cite{ZhangCR2012}. A generalized Green's function $\mathcal{G}(z,z',q;\omega)$ for this system is deduced in conjunction with the Poisson equation for the statics (the in-plane wavenumber $q\rightarrow0$ and frequency $\omega\rightarrow0$) and  dynamic response. (See Supplemental Material for details.)

Several high-index dielectrics, Al$_2$O$_3$, HfO$_2$, and TiO$_2$, are investigated; and the common medium-/low-index dielectrics, SiO$_2$ and air, are included for comparison. Table \ref{TabParameters} lists the material properties adopted or obtained in our calculation. Figure~\ref{FigGroundState} shows the near-interface ground-state electron density profile $n_{\SSS}(z)/\bar{n}$ and effective potential $V_{\text{eff}}(z)$. With increasing static permittivity and electron affinity, the potential barrier (work function) drops by as much as 2~eV, and an increasing number of electrons spill from Ag into the dielectrics. This behavior can be quantified by a characteristic spillover depth, $\zeta \equiv \int_0^{+\infty} \Dd z\ n_{\SSS-}(z)/\bar{n}$ representing the distance up to which the spilled density would extend if it had the constant bulk value $\bar{n}$ \cite{LiebschBook,PerssonPRB1985}. As shown in Table \ref{TabParameters}, $\zeta$ of HfO$_2$ or TiO$_2$ is 2 to 3 times greater than that of air, and has approached the Thomas-Fermi screening length, $l_{\SSS\text{TF}}\approx0.58$~{\AA} of Ag. The actual density tail penetrates several times further, as displayed in Fig.~\ref{FigGroundState}.

\begin{table}[hbt]
\caption{Adopted material constants of the dielectrics \cite{RobertsonRPP2006,PalikBook,GrantRMP1959,TanemuraASS2003,ZhengJPCA2005,ZhangJMC2009,ScanlonNM2013}: electronic bandgap $E_g$, electron affinity $\AlphaD$, static permittivity $\EpsDO\equiv\EpsD(\omega\rightarrow0)$, and permittivity at 532 nm optical frequency $\EpsD(\omega_{\SSS \text{532}})$. Calculated Ag-to-dielectric work function $W$ and spillover depth $\zeta$.}
\begin{center}
\begin{ruledtabular}
\begin{tabular}{c|ccccccc}
Material & $E_g$ (eV) & $\AlphaD$ (eV) & $\EpsDO$ & $\EpsD(\omega_{\SSS \text{532}})$ & $W$ (eV) & $\zeta$ (\AA) \\
\hline
Air & ``$\infty$" & 0 & 1 & 1 & 4.26 & 0.18 & \\
\hline
SiO$_2$ & 9.0 & 0.95 & 3.9 & 2.13 & 3.22 & 0.24 & \\
\hline
Al$_2$O$_3$ & 8.8 & 1.70 & 9.0 & 3.13 & 2.73 & 0.32 & \\
\hline
HfO$_2$ & 5.8 & 2.14 & 25.0 & 4.33 & 2.47 & 0.39 & \\
\hline
TiO$_2$ & 3.2 & 3.00 & 86.0 & 6.55 & 2.26 & 0.58 & \\
\end{tabular}
\end{ruledtabular}
\end{center}
\label{TabParameters}
\end{table}

\begin{figure}[htb]
\centerline{\includegraphics[scale=0.45]{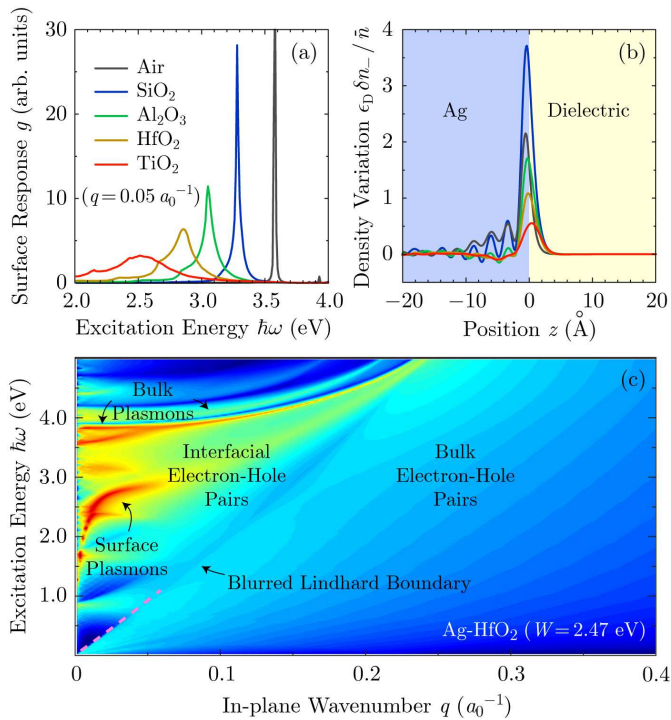} }
\caption{(a) Calculated dynamic surface response function. (b) Calculated induced density variation (scaled by $\EpsD$ for clarity) of surface plasmons at peak response energy. (c) Calculated energy-momentum loss spectrum $\mathit{\Gamma}(\bm{q},\omega)$ for Ag-HfO$_2$, plotted in logarithmic scale. Red means high loss and blue means low loss. Note that for a $100~\rB$ thin Ag slab in our model, there can be more than one bulk-plasmon curves.}
\label{FigDynamicResponseCombine}
\end{figure}

To see how quantum spillover influences the dynamic excitations, we implement a time-dependent linear-response calculation \cite{YuanPRB2006,AbajoRMP2010}. The system-mediated effective interaction between two external charge sheets at $Z$ and $Z'$ is \cite{AbajoRMP2010}
\begin{equation}
\begin{split}
\mathcal{W}(Z,Z',q;\omega) = \ & \int\Dd z\Dd z'\ \mathcal{G}(Z,z,q;\omega) \\
& \times \chi(z,z',q;\omega) \mathcal{G}(z',Z',q;\omega),
\end{split}
\end{equation}
where $\chi(z,z',q;\omega)$ is the susceptibility. (See Supplemental Material.) The surface response function,
$
g(q;\omega) \equiv  (q/2\pi)\EpsD(\omega) \IM\mathcal{W}(Z_0,Z_0,q;\omega)
$
describes the amplitude of surface excitations caused by an external charge sheet at $Z_0$, which we take to be $50~\rB$ outside the Ag \cite{LiebschBook}.
Figure \ref{FigDynamicResponseCombine}(a) gives the calculated $g(q;\omega)$ at a representative wavenumber $q=0.05~\rB^{-1}$, which amounts to a 6.6~nm wavelength, a typical length scale relevant to both optical and electronic excitations. The main peaks in Fig.~\ref{FigDynamicResponseCombine}(a) correspond to the SP excitations. As the dielectric index rises, they change from a narrow 3.57~eV resonance for air to a broad feature around 2.51~eV for TiO$_2$. Fig.~\ref{FigDynamicResponseCombine}(b) shows the induced density variation corresponding to the peak frequencies, and confirms the surface-mode profiles.

The spectral broadening (and the associated SP damping) here takes place \emph{without} phonon scattering or interband transitions (our density-functional model does not include these mechanisms). The intrinsic dissipation occurs via interfacial e-h pair production. According to Feibelman \textit{et al.} \cite{FeibelmanPSS1982,PerssonPRB1985}, the decay rate of SPs through e-h pairs is generally proportional to the imaginary part of a $d_{\SSS\perp}$ parameter, which is subsequently proportional to the spillover depth $\zeta$ through sum rules. A classical system with $\zeta\equiv 0$ cannot produce any out-of-plane e-h pairs. For metal-low-index contacts, $\zeta$ is practically too small to activate this dissipation channel. Only for the metal-high-index contacts studied here, $\zeta$ becomes appreciable and thus e-h pair production can induce significant effects.

To see the underlying physics comprehensively, we calculate the energy-momentum loss spectrum $\mathit{\Gamma}(\bm{q},\omega)$ for a high-energy electron penetrating the system \cite{AbajoRMP2010,PinesBook},
\begin{equation}
\begin{split}
\mathit{\Gamma}(\bm{q},\omega) =\ & - \mathcal{C}
 \int\Dd Z\Dd Z'\ \cos\left[ \frac{\omega}{v_{\text{in}} } (Z-Z') \right] \\
 &\times \IM \mathcal{W}(Z,Z',q;\omega) ,
\end{split}
\end{equation}
where $\mathcal{C}$ is a universal constant and $v_{\text{in}}$ is the incoming velocity of the electron (80~keV in our calculation and experiment). For a conventional Ag-vacuum contact (the work function $W=4.26$~eV), the spectrum should contain a clear Lindhard boundary between the collective excitations (surface and bulk plamsons) and the bulk e-h pair production \cite{GiulianiBook}. Unless the Landau damping comes in at extremely large momenta, SPs cannot decay into \emph{bulk} e-h pairs without violating energy-momentum conservation (see Supplemental Material for details). However, our calculated spectrum Fig.~\ref{FigDynamicResponseCombine}(c) for an Ag-HfO$_2$  contact ($W=2.47$~eV) exhibits a thoroughly blurred region between the collective excitations and the bulk e-h pair production. This is the region of \emph {interfacial} e-h pair production. For a metal-dielectric contact of a small work function (low barrier), electrons can actively move out-of-plane irrespective of how small the in-plane $q$ is. SPs can spontaneously decay into $\emph{interfacial}$ e-h pairs without violating energy-momentum conservation.
This manifests the importance of ground-state properties to the optical-frequency behaviors in such systems.

\begin{figure}[hbt]
\centerline{\includegraphics[scale=0.8]{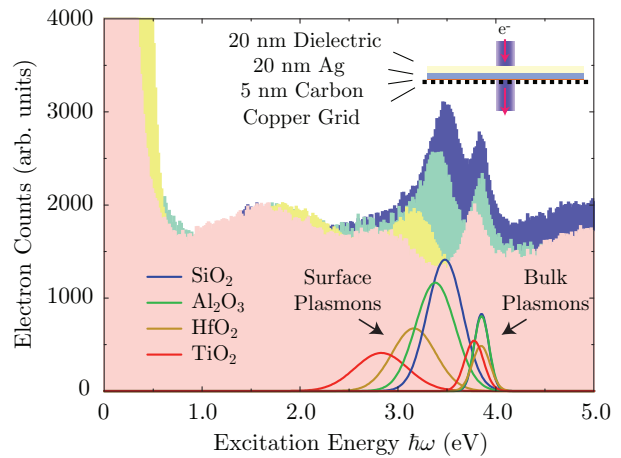}}
\caption{Measured electron energy-loss spectra. The colored column plots display the electron counts. The colored curves show the peaks via the standard multi-Lorentzian fitting (after subtracting the zero-loss peak).} \label{FigEELS}
\end{figure}

Looking for experimental evidence, we perform electron energy-loss spectroscopy (EELS) measurement (aberration corrected Zeiss Libra transmission electron microscope (TEM)). The acceleration voltage of the microscope is 80 kV, which permits an excellent energy resolution of 120 meV (FWHM of the zero-loss peak). We e-beam evaporate 20~nm Ag and 20~nm varied dielectrics sequentially onto carbon-supported TEM grids. Figure \ref{FigEELS} shows the obtained energy-loss spectra. The broad peaks in the range of 1.5 to 1.8~eV for all dielectrics come from the contact between 5~nm amorphous carbon and Ag \cite{RobertsonPRB1987}. These peaks are not of interest. The sharp peaks around 3.8~eV for all dielectrics are the bulk-plasmon resonances inside Ag \cite{AndersenPRB2012}. The varied peaks consistently moving from about 3.5~eV for SiO$_2$ to about 2.8~eV for TiO$_2$ are the SP resonances at the Ag-dielectric interfaces. There is a clear trend that, with increasing index of the dielectrics, the SP peak-width gradually broadens. This trend signifies an enhanced damping and shortened lifetime of the SPs, in agreement with our theoretical prediction in Fig.~\ref{FigDynamicResponseCombine}. By comparison, the bulk-plasmon resonances, which decay primarily through phonon scattering (the Drude loss) and bulk e-h pair production (only at extremely short wavelengths), are much less affected by the varied dielectrics. This observation testifies to the dominant role of interfacial e-h pair production, as opposed to phonon scattering, in the SP absorption at Ag-high-index interfaces.

In order to directly connect to optical measurement, we further develop a mesoscopic formulation to quantify the crossover from the phonon-dominant regime to the e-h-pair-dominant regime. The $p$-wave reflection coefficient at a metal-dielectric interface allowing quantum spillover can be derived by generalizing the original formulation by Feibelman \textit{et al.} \cite{LiebschBook,FeibelmanPSS1982,PerssonPRB1985,KempaPRB1986},
\begin{align}
r^{(p)} (q,\omega) &=\frac{\DS \frac{\kappaD}{\EpsD}-\frac{\kappaM}{\EpsM} + \frac{\EpsM-\EpsD}{\EpsM \EpsD} \left[\dperp q^2 + \dpara \kappaD\kappaM \right]}{\DS \frac{\kappaD}{\EpsD} + \frac{\kappaM}{\EpsM} - \frac{\EpsM-\EpsD}{\EpsM \EpsD} \left[\dperp q^2 - \dpara \kappaD\kappaM \right]}.\label{EqnRP}
\end{align}
Here, $\EpsM(\omega)=\EpsB(\omega)-\OmegaP^2/(\omega^2+\Ii\GammaP\omega) $ is the Lorentz-Drude permittivity of bulk Ag (see Supplemental Material), $\kappaD(q,\omega)=\sqrt{q^2-\EpsD(\omega)\omega^2/c^2}$, $\kappaM(q,\omega)=\sqrt{q^2-\EpsM(\omega)\omega^2/c^2}$ are evanescent wavenumbers. $\dperp$ and $\dpara$ are two complex-valued parameters, associated with the long-wavelength electron-density oscillation near the interface. If $\dperp$ and $\dpara$ both vanish, then Eq.~(\ref{EqnRP}) is reduced to the standard $p$-wave Fresnel reflection coefficient that embodies classical SP resonance when the denominator $\frac{\kappaD}{\EpsD} + \frac{\kappaM}{\EpsM}$ is near zero. $\dpara$ typically vanishes for uncharged surfaces. The real part of $\dperp$ gives the nonlocal correction to the SP dispersion. The imaginary part gives the more important e-h pair loss strength. Our density-functional calculation gives $\IM\dperp\approx1.1$~{\AA} for SiO$_2$, $1.5$~{\AA} for Al$_2$O$_3$, $2.1$~{\AA} for HfO$_2$ and $3.0$~{\AA} for TiO$_2$ \cite{LiebschBook}. Note that $\IM \dperp$ is larger than the ground-state spillover depth $\zeta$ listed in Table~\ref{TabParameters}.

\begin{figure}[hbt]
\centerline{\includegraphics[scale=0.8]{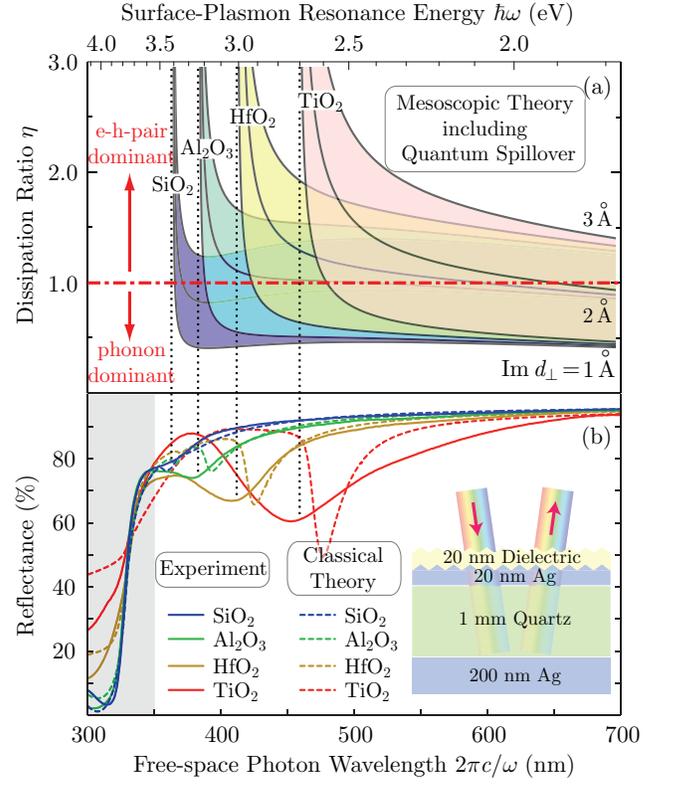}}
\caption{(a) Calculated dissipation ratio of the e-h-pair loss versus the phonon loss around the conventional SP dispersion curve. (b) Measured ultraviolet-visible reflection spectra, and calculated reflection spectra using classical roughness theory. The shaded region ($\lambda<350$~nm) is where silver approaches bulk-plasmon absorption and can have larger uncertainties. } \label{FigDissipationRatio}
\end{figure}

The phonon (Drude) loss of Ag is small compared with visible frequencies, so a linear expansion to $\GammaP/\omega$ in Eq.~(\ref{EqnRP}) around the conventional SP dispersion curve $\omega(q) $ (solved from $q^2= [\BarEpsM\EpsD/(\BarEpsM+\EpsD)]\omega^2/c^2$) is legitimated. Here $\BarEpsM(\omega)$ is the $\GammaP\rightarrow0$ limit of $\EpsM(\omega)$. We can determine a dissipation ratio $\eta[\omega(q)]$ (see Supplemental Material) that evaluates the strength of e-h pair loss versus the phonon loss at each SP-resonance frequency \emph{along} the dispersion curve $\omega(q)$,
\begin{equation}
\eta[\omega(q)] = \frac{\DS 2 |\BarEpsM(\omega)|^2 }{\sqrt{|\BarEpsM(\omega)|-\EpsD(\omega)}} \frac{\omega^4}{\OmegaP^2} \frac{\IM\dperp}{\GammaP c}.\label{EqnLossCompare}
\end{equation}
In Fig.~\ref{FigDissipationRatio}(a), we plot $\eta$ as a function of $\omega$ for all the dielectrics, with the value of $\GammaP$ fitted from real Ag, and with $\IM \dperp$ varying from $1$~{\AA} to $3$~{\AA}. The red dashed-dotted line separates the phonon-dominant regime ($\eta(\omega)<1$) from the e-h-pair-dominant regime ($\eta(\omega)>1$). The conventional SP-dispersion asymptotically approaches the frequency where $|\BarEpsM(\omega)|=\EpsD(\omega)$ (the black dashed lines in Fig.~\ref{FigDissipationRatio}(a)). At this frequency (corresponding to very large $q$), the e-h pair loss always dominates, even for a medium-index dielectric like SiO$_2$. However, towards smaller frequencies (longer $2\pi c/\omega$ but still fairly large $q$), Ag-SiO$_2$ with $\IM\dperp\approx1.1$~{\AA} immediately drops to the Drude-dominant regime. Ag-Al$_2$O$_3$ with $\IM\dperp\approx1.5$~{\AA} is at the onset where e-h pair loss starts to take effect in a wider range up to $2\pi c/\omega\approx 400$~nm or so. Ag-HfO$_2$ with $\IM\dperp\approx2.1$~{\AA} has an even wider e-h-pair-dominant range beyond 500~nm. Ag-TiO$_2$ with $\IM\dperp\approx3.0$~{\AA} is fully e-h-pair-dominant covering the entire visible spectrum. This result suggests possible observation, via optical approach, of strong and broadened resonant absorption around the large-momentum SP frequency (around the black dotted-lines in Fig.~\ref{FigDissipationRatio}(a)), especially for Ag-high-index contacts.

We perform ultraviolet-visible (UV-Vis) spectrophotometry measurement (Varian Cary 500). We e-beam evaporate 20~nm Ag films onto ultra-smooth (roughness $\delta<3$~{\AA}) $z$-cut single-crystal quartz substrates, and conformally coat 20~nm varied dielectrics on the top via atomic layer deposition (ALD). The 20~nm Ag on the dielectric side has a roughness $\delta\approx2$~nm and a correlation length $\sigma\approx35$~nm measured by atomic-force microscopy (AFM). This slightly roughened surface provides the desired large momenta to excite SPs under normally incident light without a need of  prism coupling \cite{RaetherBook,ElsonPRB1971,KrogerZPhys1970,KretschmannJOSA1975}. For convenience, we deposit additional 200~nm Ag on the bottom of the substrates to block the transmission, so we can measure the reflection spectrum from the top (refer to the inset of Fig.~\ref{FigDissipationRatio}(b)). In the wavelength range of 350 to 700~nm, the quartz substrates are completely transparent and the 200~nm Ag on the bottom serves as an ideal mirror. Any drop on the reflection spectrum is due to the absorption from the 20~nm dielectric and 20~nm Ag on the top (diffusive scattering is negligible at this roughness). The solid curves in Fig.~\ref{FigDissipationRatio}(b) show the observed resonant absorption around the large-momentum SP-frequency of each dielectric. The reflection dip is not prominent for SiO$_2$, but continuously intensifies and widens from Al$_2$O$_3$ to TiO$_2$, as anticipated from Fig.~\ref{FigDissipationRatio}(a). The observed absorption is not caused by interband transitions (dielectric loss) inside the dielectrics, as we have done separate transmission measurement for the dielectrics on quartz alone (no Ag) to confirm their lossless nature in the wavelength range of interest (see Supplemental Material). It is not solely due to the phonon-scattering (Drude loss) in Ag either, as we have carried out a classical transfer-matrix calculation combined with the statistical-roughness theory following Kretschmann \textit{et al.} \cite{KrogerZPhys1970,KretschmannJOSA1975}. This calculation encloses all the Drude absorption at the interface and bulk of Ag. The dashed curves in Fig.~\ref{FigDissipationRatio}(b) show the reflection dips from the classical theory. They are sharper and located at longer wavelengths (determined by the $\sigma\approx35$~nm correlation length), in contrast to the experimental curves. Only SiO$_2$ exhibits a similarity between the experimental and (classical) theoretical curves. All the high-index dielectrics show large discrepancies. Based on our systematic theoretical development above, the e-h-pair-dominant loss is the most probable reason for the observed  strong and broadened absorption of short-wavelength SPs at Ag-high-index interfaces.

To summarize, we find that high-index dielectrics contacting with silver can exhibit enhanced surface-plasmon absorption due to the quantum-spillover supported interfacial electron-hole pair production. The quantum-electronic control of the static dielectric environment to the optical excitations on metal surfaces can bring on new applications in nanoscale light confinement and new insights in surface-plasmon to hot-electron conversion.

\begin{acknowledgements}
D. J., Q. H., Y. Y., and N. X. F. acknowledge the financial support by the NSF (Award No. CMMI-1120724) and AFOSR MURI (Award No. FA9550-12-1-0488). D. N. acknowledges support by the NSF (Grant CHE-1112500). F. v. C. acknowledges support by the STC Center for Integrated Quantum Materials and NSF (Grant No. DMR-1231319). DJ wish to thank Patrick A. Lee, Fan Wang, Sang Hoon Nam, Miguel A. M\'{e}ndez Polanco, and Alexie M. Kolpak for helpful discussions.
\end{acknowledgements}

\end{document}